\documentclass[conference]{IEEEtran}
\IEEEoverridecommandlockouts
\usepackage{cite}
\usepackage{amsmath,amssymb,amsfonts}
\usepackage{algorithmic}
\usepackage{graphicx}
\usepackage{textcomp}
\usepackage{cite}
\usepackage{algorithmic}
\usepackage{graphicx}
\usepackage{textcomp}
\usepackage[dvipsnames]{xcolor}
\usepackage{paralist}
\usepackage[inline]{enumitem}
\usepackage{caption}
\usepackage{multirow}
\usepackage{booktabs}
\usepackage{makecell}
\usepackage{tabu, booktabs}
\usepackage{float}
\usepackage{subcaption}
\usepackage{pifont}
\usepackage{comment}
\usepackage{todonotes}
\usepackage{hyperref}
\usepackage{csquotes}
\usepackage{multirow}
\usepackage{booktabs}
\usepackage{graphicx}
\usepackage{epstopdf}
\usepackage{color}
\usepackage{gnuplottex}
\usepackage{graphicx}
\usepackage{epstopdf}
\usepackage{pgf}

\newcommand\myshade{85}

\definecolor{magma_darker}{HTML}{fdc38a}
\definecolor{magma_dark}{HTML}{e15666}
\definecolor{magma_lighter}{HTML}{1f0c43}

\definecolor{cvpr_link}{HTML}{00ffff}
\definecolor{cvpr_cite}{HTML}{00ff00}
\definecolor{cvpr_file}{HTML}{ff0000}

\definecolor{royalblue}{RGB}{65, 105, 225}
\definecolor{darkgreen}{RGB}{0, 100, 0}
\definecolor{darkred}{RGB}{139, 0, 0}
\definecolor{darkviolet}{RGB}{148, 0, 211}

\colorlet{linkColor}{violet}
\colorlet{citeColor}{YellowOrange}
\colorlet{urlColor}{Emerald}

\hypersetup{
  linkcolor  = darkviolet!\myshade!black,
  citecolor  = citeColor!\myshade!black,
  urlcolor   = urlColor!\myshade!black,
  filecolor  = darkgreen!\myshade!black,
  colorlinks = true,
}
\usepackage[a4paper, total={184mm,239mm}]{geometry}
\def\BibTeX{{\rm B\kern-.05em{\sc i\kern-.025em b}\kern-.08em
    T\kern-.1667em\lower.7ex\hbox{E}\kern-.125emX}}

\ExplSyntaxOn
\DeclareExpandableDocumentCommand{\convertlen}{ O{cm} m }
 {
  \dim_to_decimal_in_unit:nn { #2 } { 1 #1 } cm
 }
\ExplSyntaxOff
\usepackage{gnuplottex}
\usepackage{tikz}
\usepackage{mathtools}

\usetikzlibrary{matrix,calc,positioning,arrows.meta}

\begin{document}

\title{Late Breaking Results: Leveraging Approximate Computing for Carbon-Aware DNN Accelerators}

\author{
\IEEEauthorblockN{
Aikaterini Maria Panteleaki\IEEEauthorrefmark{1},
Konstantinos Balaskas\IEEEauthorrefmark{2},
Georgios Zervakis\IEEEauthorrefmark{2},
Hussam Amrouch\IEEEauthorrefmark{3},
Iraklis Anagnostopoulos\IEEEauthorrefmark{1}
}
\IEEEauthorblockA{
\IEEEauthorrefmark{1}Southern Illinois University Carbondale, 
\IEEEauthorrefmark{2}University of Patras, 
\IEEEauthorrefmark{3}Technical University of Munich
}
}

\maketitle
\begin{abstract}
The rapid growth of Machine Learning (ML) has increased demand for DNN hardware accelerators, but their embodied carbon footprint poses significant environmental challenges. This paper leverages approximate computing to design sustainable accelerators by minimizing the Carbon Delay Product (CDP). Using gate-level pruning and precision scaling, we generate area-aware approximate multipliers and optimize the accelerator design with a genetic algorithm. Results demonstrate reduced embodied carbon while meeting performance and accuracy requirements.
\end{abstract}



\vspace{-5pt}
\section{Introduction}\label{sec:introduction}
\vspace{-3pt}
The rapid growth of machine learning (ML) has driven advances in computing, with specialized hardware accelerators enhancing the efficiency of Deep Neural Networks (DNNs). However, this progress comes with a significant environmental cost, as the embodied carbon footprint from the manufacturing of these accelerators remains largely unexplored. Recent studies~\cite{gupta2021chasing,panteleaki2024carbon} highlight that embodied carbon now surpasses operational emissions as a dominant factor in the environmental impact of ML systems, particularly in edge-based applications.

Designing DNN hardware accelerators is challenging due to the wide range of possible hardware configurations and mappings. Key decisions, such as determining the number of Processing Elements (PEs) and setting up local and global memory configurations, greatly affect the accelerator's performance. However, previous works~\cite{gupta2022act} have shown that such accelerators are often overdesigned, providing more performance than necessary for edge applications, and significantly increasing their embodied carbon footprint at the same time. Relaxing performance requirements offers a promising solution that allows the development of carbon-aware designs, that better balance performance and sustainability.


Moreover, DNNs are inherently resilient to computational errors in arithmetic operations, making them an ideal candidate for leveraging approximate computing to reduce embodied carbon. By introducing approximate arithmetic units, which require fewer transistors and have a smaller hardware footprint, it is possible to significantly lower the embodied carbon of DNN accelerators. These approximate units not only reduce the area required for computation, but also free up design space for optimizing memory configurations. Despite the potential benefits, no prior work has explored the use of approximate computing as a means to address the embodied carbon emissions of DNN accelerators. 

In this work, we investigate how relaxed constraints along with approximate computing can be systematically applied to balance embodied carbon footprint and performance for DNN inference accelerators.
\vspace{-3pt}
\section{Methodology}\label{sec:methodology}
\vspace{-3pt}
\begin{figure}
    \centering
    \resizebox{0.35\textwidth}{!}{\includegraphics[clip]{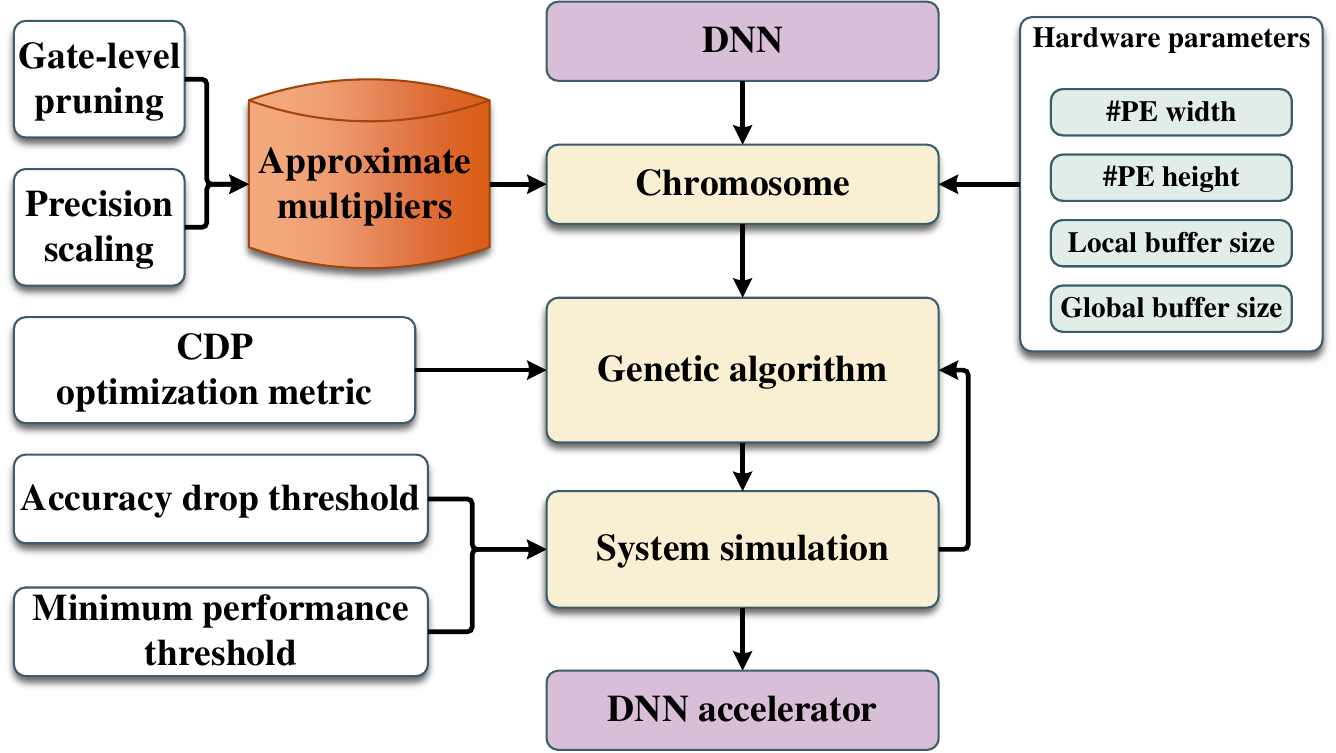}}
    \caption{Overview of the proposed methodology}
    \vspace{-10pt}
    \label{fig:overview}
\end{figure}

The primary objective of our methodology is to design an approximate DNN accelerator and determine its corresponding mapping to optimize the Carbon Delay Product (CDP). CDP is a comprehensive metric that integrates performance and the embodied carbon footprint, offering a holistic assessment of the trade-offs between sustainability and efficiency.
By focusing on minimizing the CDP, we aim to design hardware that achieves a balance between performance and carbon emissions, ensuring suitability for edge-based systems.


The embodied carbon estimation of an accelerator, while related to chip area, primarily depends on the die manufacturing process, which incorporates several technology and fab-specific factors beyond area alone. Key contributors include the fabrication facility's attributes, such as its power consumption and the carbon intensity of its electricity grid. Additionally, the technology node used in the fabrication process significantly impacts scaling trends and yield results. For a monolithic DNN accelerator die, the embodied carbon footprint is calculated based on emissions produced during the manufacturing of its logic chip area, using a specified technology node \cite{sudarshan2024eco}. The total embodied carbon of a chip comprises two main components: the product of the Carbon Footprint Per unit Area (CFPA) of the die and its area (A$_\text{die}$), and the product of the CFPA of Silicon (CFPA$_\text{Si}$) and the wasted area of the silicon wafer (A$_\text{wasted}$) during fabrication, as shown in Eq. \ref{eq:embodied}. The CFPA depends on factors such as the Carbon Intensity of the fabrication facility (CI$_\text{fab}$), the Energy consumed per unit Area during manufacturing (EPA), the greenhouse gas emissions (C$_\text{gas}$), the carbon impact of raw material procurement (C$_\text{material}$), and the yield (Y) of the fabrication process.

\vspace{-6pt}
\begin{equation}
    C_{\text{embodied}} = \text{CFPA} \times A_{\text{die}} + \text{CFPA}_{\text{Si}} \times A_{\text{wasted}}
     \label{eq:embodied}
\end{equation}
\vspace{-6pt}
\begin{equation}
    \text{CFPA} =  \frac{\text{CI}_\text{fab} \times \text{EPA} + \text{C}_\text{gas} + \text{C}_\text{material}}{\text{Y}}
     \label{eq:cfpa}
\end{equation}

To optimize embodied carbon, while maintaining computational efficiency, we start by generating area-aware approximate multipliers for the MAC units. To achieve this, we apply gate-level pruning and precision scaling approximation techniques to modify the netlist structure or the connections between its gates, effectively reducing the circuit area \cite{9804695}. These approximations are guided by a multi-objective optimization algorithm that explores the design space to identify near-Pareto-optimal solutions with minimal functional error. The resulting approximate multipliers not only lower the embodied carbon footprint by reducing the required hardware area but also maintain the computational accuracy needed for error-resilient DNN tasks.

In the second step, we integrate these approximate multipliers into exploring hardware configurations and mappings for the accelerator. This involves optimizing key characteristics such as the width and height of the accelerator (number of Processing Elements), local register file sizes, and global buffer capacity. Mapping parameters, including tiling strategies, execution order, and levels of parallelism, are also considered. 
To navigate this vast design space, we employ a genetic algorithm, with CDP metric as fitness function, to select the Pareto-optimal approximate multipliers from step one and identify the most efficient DNN topology. The optimization process is constrained by thresholds for accuracy drop and performance, measured in inferences per second, ensuring that the design meets the realistic requirements of edge systems. This approach addresses the overdesign issue observed in previous accelerators, resulting in a more sustainable and efficient solution.


\vspace{-3pt}
\section{Evaluation}\label{sec:evaluation}
\vspace{-5pt}
As a baseline for our design exploration, we use the NVDLA architecture paradigm, which include MAC arrays ranging from 64 to 2048 PEs in powers of $2$. The sizes of the local and global convolution buffers scale proportionally with the dimensions of the MAC arrays, as specified by NVIDIA \cite{nvdla2017}. To evaluate these configurations, we employ two specialized tools: the nn-dataflow~\cite{gao2019tangram}, to estimate DNN workload performance, and ApproxTrain~\cite{gong2023approxtrain}, to calculate the accuracy impact of the approximate multipliers.

\begin{figure}[t]
    \vspace{-10pt}
    \centering
    \begin{minipage}[t]{0.5\linewidth}
        \centering
        \vspace{-2pt}
        \resizebox{\linewidth}{!}{\includegraphics{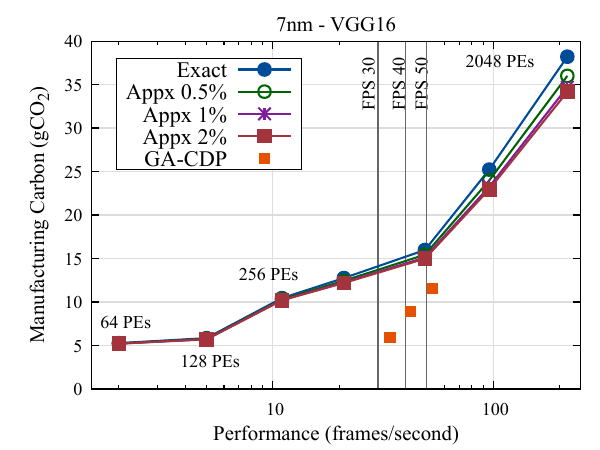}}
        \label{fig:trend}
    \end{minipage}
    \hfill
    \begin{minipage}[t]{0.45\linewidth}
    \centering
    \vspace{15pt}
    \resizebox{\linewidth}{!}{%
    \begin{tabular}{|c|c|c|c|c|}
    \hline
    \multicolumn{5}{|c|}{Carbon Footprint Reduction (\%)} \\
    \hline
    \multirow{2}{*}{Technology} & \multirow{2}{*}{Type} & \multicolumn{3}{c|}{Accuracy Drop } \\
    \cline{3-5}
    (nm) &  & 0.5\% & 1.0\% & 2.0\% \\
    \hline
    \multirow{2}{*}{7} & Avg & 2.83 & 4.49 & 5.17 \\
                       & Peak & 5.78 & 9.18 & 10.56 \\
    \hline
    \multirow{2}{*}{14} & Avg & 5.58 & 6.90 & 8.02 \\
                        & Peak & 8.87 & 10.98 & \textbf{12.75} \\
    \hline
    \multirow{2}{*}{28} & Avg & 3.33 & 5.71 & 8.44 \\
                        & Peak & 4.60 & 7.87 & 11.65 \\
    \hline
    \end{tabular}
    }
    \label{fig:carbon_table}
    \end{minipage}
    \vspace{-15pt}
    \caption{Embodied Carbon trends for VGG16 across different accuracy and performance levels.}
    \label{fig:combined}
\end{figure}

Figure \ref{fig:combined} illustrates the trade-off between embodied carbon and performance for DNN accelerators running VGG16 at the 7nm technology node. The configurations for the exact (baseline) accelerator show exponential carbon increase as the architecture becomes more compute-intensive. By incorporating approximate units only, while keeping the architecture unchanged (same number of PEs and memory), we achieved a $5\%$ reduction in embodied carbon. In our experiments, we tested approximate units that resulted in accuracy losses of up to 0.5\%, 1.0\%, and 2.0\% (Appx in legend). Similar trends appeared at 14nm and 28nm, as shown in the corresponding table, with gains of up to $12.75\%$. 
However, the frames per second (FPS) achieved by large accelerators often far exceed the requirements for edge applications~\cite{gupta2022act}. To address this, we applied realistic performance thresholds of 30, 40, and 50 FPS and utilized our genetic algorithm with approximate multipliers. This approach significantly reduced the embodied carbon footprint, achieving reductions of up to 50\%. Each point on the plot (GA-CDP in legend) represents an accelerator design that optimizes the CDP while meeting the specified performance thresholds (30, 40, and 50 FPS).

\begin{figure}
    \vspace{-13pt}
    \centering
    \resizebox{0.48\textwidth}{!}{\includegraphics[clip]{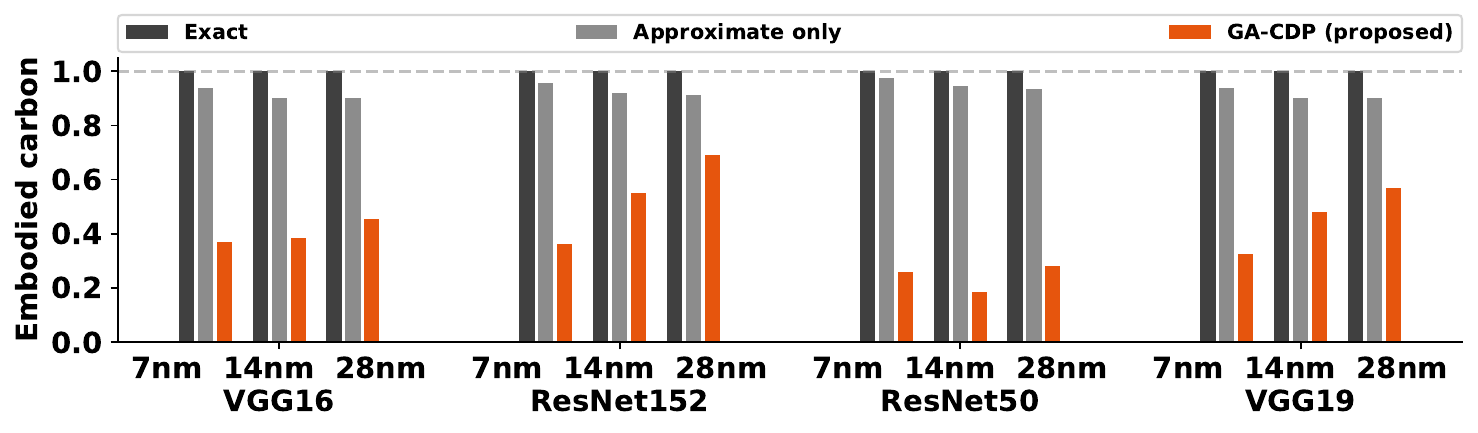}}
    \vspace{-5pt}
    \caption{Embodied carbon comparison across DNN models \\ (Normalized to exact implementation for each technology node)}
    \vspace{-15pt}
    \label{fig:overview2}
\end{figure}

To validate our methodology, we evaluated VGG16, VGG19, ResNet50, and ResNet152, on ImageNET subset, across $7$, $14$, and $28$ nm nodes. Figure \ref{fig:overview2} compares three designs: the exact baseline meeting a $30$ FPS threshold, an approximate version using area-efficient multipliers with up to $2.0\%$ accuracy drop, and our proposed solution (GA-CDP). While approximation alone reduces embodied carbon, our methodoogy further optimizes the design through minimal architecture and efficient multipliers. Results show significant reductions in embodied carbon across all networks and nodes, with up to $65\%$ savings for VGG16 and $30\%$–$70\%$ for others, proving that our approach creates sustainable, performance-compliant designs.

\vspace{-3pt}
\section{Conclusion}\label{sec:conclusion}
\vspace{-3pt}
We proposed a carbon-aware DNN accelerator design using approximate computing and architecture exploration. Our approach achieves up to $70\%$ lower carbon footprint with minimal accuracy loss, meeting performance requirements across various models and technology nodes.

\section*{Acknowledgments}\label{sec:acknowledgements}
This work has been supported by grant NSF CCF 2324854.


\footnotesize

\end{document}